# Liquid Metal-Exfoliated SnO$_2$-Based Mixed-dimensional Heterostructures for Visible-to-Near-Infrared Photodetection

Shimul Kanti Nath[1,2]*, Nitu Syed[3,4]*, Wenwu Pan[2], Yang Yu[5], Dawei Liu[5,6], Michael P. Nielsen[1], Jodie Yuwono[7], Priyank Kumar[8], Yan Zhu[1], David L. Cortie[9], Chung K. Nguyen[4], Lan Fu[5], Ann Roberts[3], Lorenzo Faraone[2], Nicholas J. Ekins-Daukes[1], and Wen Lei[2]

[1]School of Photovoltaic and Renewable Energy Engineering, University of New South Wales (UNSW Sydney), Kensington NSW 2052, Australia
[2]Australian Research Council Centre of Excellence for Transformative Meta-Optical Systems, Department of Electrical, Electronic and Computer Engineering, The University of Western Australia, 35 Stirling Highway, Perth, WA 6009, Australia
[3]Australian Research Council Centre of Excellence for Transformative Meta-Optical Systems, School of Physics, University of Melbourne, Victoria 3010, Australia
[4]School of Engineering, RMIT University, Melbourne, VIC, 3001 Australia
[5]Australian Research Council Centre of Excellence for Transformative Meta-Optical Systems, Department of Electronic Materials Engineering, Research School of Physics, The Australian National University, Canberra ACT 2601, Australia
[6]Institute of Applied Physics, Abbe Center of Photonics, Friedrich Schiller University Jena, Albert-Einstein-Straße 15, Jena, 07745, Germany
[7]School of Chemical Engineering, The University of Adelaide, Adelaide, SA 5005, Australia
[8]School of Chemical Engineering, University of New South Wales (UNSW Sydney), Kensington NSW 2052, Australia
[9]Australian Centre for Neutron Scattering, Australian Nuclear Science and Technology Organisation, Lucas Heights, NSW 2234, Australia
*Corresponding authors
   *Shimul_kanti.nath@unsw.edu.au
   *nitu.syed@unimelb.edu.au

**Abstract:** Ultra-thin two-dimensional (2D) materials have gained significant attention for making next-generation optoelectronic devices. Here, we report a large-area heterojunction photodetector fabricated using a liquid metal-printed 2D SnO$_2$ layer transferred onto CdTe thin films. The resulting device demonstrates efficient broadband light sensing from visible to near-infrared wavelengths, with enhanced detectivity and faster photo response than bare CdTe photodetectors. Significantly, the device shows a nearly 10$^5$-fold increase in current than the dark current level when illuminated with a 780 nm laser and achieves a specific detectivity of around 10$^{12}$ Jones, nearly two orders of magnitude higher than a device with pure CdTe thin film. Additionally, temperature-dependent optoelectronic testing shows that the device maintains a stable response up to 140 °C and generates distinctive photocurrent at temperatures up to 80°C, demonstrating its thermal stability. Using band structure analysis, density functional theory (DFT) calculations, and photocurrent mapping, the formation of a *p-n* junction is indicated, contributing to the enhanced photo response attributed to the efficient carrier separation by the built-in potential in the hetero-junction and the superior electron mobility of 2D SnO$_2$. Our results highlight the effectiveness of integrating liquid metal-exfoliated 2D materials for enhanced photodetector performance.

**1. Introduction**: Two-dimensional (2D) materials have attracted significant recent attention for their use in optoelectronics owing to their extraordinary properties emerging from ultra-thin, atomic-scale architecture [1-3]. In particular, 2D layers of metal oxides have emerged as promising candidates for their use in next-generation photodetectors [4] which have applications in a variety of areas, including telecommunications, scientific instrumentation, astronomy, surveillance, and defense [5-7]. However, the nanoscale thickness of these materials poses a significant limitation, as it permits the absorption of only a minimal portion of the incident light, thereby constraining their efficacy in light-driven applications, particularly within the visible to near-infrared (NIR) spectral range [8]. Developing innovative strategies that enhance light-matter interactions is crucial to harness their potential fully. Constructing mixed-dimensional (2D/3D) heterostructures is of particular interest because they provide several advantages, including enhanced light absorption, a broad spectral range (UV to IR), formation of built-in potential, which assists in separating photo-generated carriers leading to fast photo-response, and extended functionality such as self-powered photodetection. These attributes are useful in environments that require ultralow power consumption, a small device footprint, and wireless operation [9, 10]. Therefore, significant efforts have been made to develop heterostructures incorporating new 2D materials with high sensitivity, wide detection ranges, rapid response times, and ease of fabrication [11-13].

To date, the majority of 2D/3D p-n heterojunction-based photodetectors have concentrated on combinations such as silicon with graphene [14] or transition metal dichalcogenides (TMDCs), including MoS$_2$ [15, 16], while more recently, a limited number of studies have investigated graphene paired with GaN [17]. Importantly, heterojunctions involving non-layered 2D materials combined with 3D semiconductors remain unexplored, highlighting a promising and under-investigated area of research. The dangling bonds on the surface of non-layered 2D metal oxides exhibit high chemical activity, providing enhanced opportunities to tailor their chemical reactivity with other materials [18].

The growth and characterization of 2D material-based heterostructure devices are challenging due to the difficulty of growing large-area 2D layers on appropriate substrates. Liquid metal-based synthesis is a recently developed technique that offers an effective means of realizing various 2D metal oxides and compounds due to the relative simplicity of the synthesis process and the ability to create large, ultra-thin materials with minimal imperfections [19-21]. Due to the high surface tension, the surface of liquid metals is often described as atomically flat, providing an optimal environment for creating ultrathin 2D materials following the Cabrera–Mott oxidation process [19, 20, 22]. Moreover, liquid metal-based printing techniques facilitate synthesizing a wide range of materials, including naturally stratified, non-layered compounds, significantly expanding the available family of 2D materials. However, incorporating liquid metal-derived semiconducting oxides as a mixed heterostructure platform has rarely been reported and has remained an active area of research. Such a hybrid system could offer tuneable band alignments, exceptional optical properties, and atomically smooth interfaces, enabling efficient charge dynamics and broadband light responsiveness.

Tin oxide ($SnO_2$) has emerged as a critical material in various optoelectronic and sensing applications, including solar cells (as an electron transport layer) [23], ultraviolet (UV) photodetectors [24], and gas and alcohol sensors [25-27]. This versatility is attributed to its desirable properties, such as high chemical stability, a wide bandgap (3.4 to 4 eV), and superior electron mobility (~ 192 $cm^2V^{-1}s^{-1}$) [28-31]. On the other hand, Cadmium telluride (CdTe) is also a technologically important material that exhibits promising optoelectronic properties, including a direct bandgap of approximately 1.5 eV and a high optical absorption coefficient that enables effective light absorption above the bandgap, making it an excellent candidate for photodetector and solar cell applications [32-35]. The favorable band alignment and exceptional electron transport properties of $SnO_2$ have inspired a growing interest in integrating $SnO_2$ with CdTe technology [36, 37].

This study uses liquid metal synthesis to demonstrate a heterostructure photodetector utilizing 2D $SnO_2$ layers transferred onto bulk CdTe thin films. This mixed-dimensional photodetector demonstrates enhanced broadband light sensing and can generate photocurrents without external bias, delivering superior detectivity and faster photo response than conventional CdTe photodetectors. These advances highlight the significant benefit and potential of incorporating liquid metal-exfoliated ultrathin materials into next-generation nano optoelectronic devices.

## 2. Results and Discussion
## 2.1 Material Synthesis and Characterisation

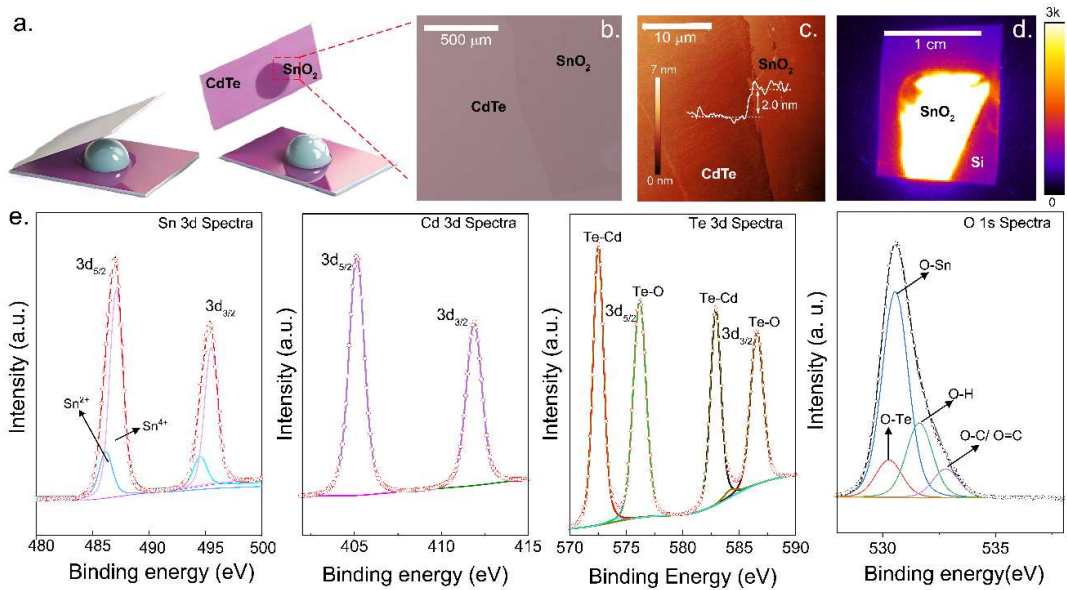

Fig. 1: (a) Schematic of the liquid metal synthesis technique, (b) optical image showing the exfoliated $SnO_2$ onto the CdTe thin films, (c) AFM image confirming the thickness of the as-deposited 2D SnO₂ layer, (d) photoluminescence image of a 2D SnO₂ transferred onto a Si substrate showing the overall cm-scale growth of the 2D material and (e) XPS spectra of SnO₂/CdTe heterostructure

Fig. 1(a) shows a schematic of the liquid metal synthesis technique used to produce centimetre-scale 2D SnO₂ layers [38]. This single-step liquid metal printing technique transfers ultrathin tin oxides onto different substrates. The process involves exfoliating 2D SnO$_x$ from freshly preconditioned tin melted at 280 °C in an ambient atmosphere (further details are given in the experimental section). Under these conditions, the resulting SnO$_x$ comprises a mixture of tin monoxide (SnO) and tin dioxide (SnO₂) [27, 38]. Post-transfer annealing in ambient air was performed to promote the SnO₂ phase, yielding near-uniform 2D nanosheets, as shown in Figs. 1(b–d). An optical image of the 2D layer transferred onto a CdTe thin film of ~ 5 μm thickness is shown in Fig. 1(b), and an AFM image of the structure is shown in Fig. 1(c), identifying a SnO₂ thickness of ~ 2 nm with minimal surface roughness. Fig. 1(d) shows a photoluminescence image of a 2D SnO₂ layer produced by the same process but transferred onto a bare Si substrate, demonstrating the compatibility of the synthesis method across various substrates. Here, SnO₂ on Si serves as a surface passivation layer, enabling visualization of the quality and yield of the 2D layer obtained through this technique.

The chemical states of the SnO₂/CdTe interface were analyzed using X-ray photoelectron spectroscopy (XPS), with the resulting spectra for Sn 3d, Cd 3d, Te 3d, and O 1s shown in Fig. 1(e). The XPS analysis reveals the presence of $Sn^{4+}$ and $Sn^{2+}$, with the dominant response resulting from the presence of the SnO₂ phase. Note that the device fabrication using photolithography involved subjecting the materials to multiple brief annealing steps in air, water rinsing, and atmospheric storage with no encapsulation before electrical measurements. This processing potentially affects Sn oxidation states, with SnO₂ likely prevailing under these conditions [27, 39, 40]. This is also supported by Raman measurements (Fig. S1) and thermodynamic considerations (see supporting information). Additionally, the XPS data indicate Te oxidation (formation of TeO₂ and TeO$_x$ [41]) at the surface, consistent with observations seen in Te-terminated CdTe grown by molecular beam epitaxy (see supporting information Fig. S2). The O 1s spectrum shows peaks associated with various oxygen species for the SnO₂-CdTe heterostructure.

## 2.2 Photodetection performance

A two-terminal photodetector partly incorporating the uncovered CdTe thin film and the SnO₂/CdTe heterostructure, as shown in Fig. 2(a), was fabricated. The device response was investigated under illumination with lasers of different wavelengths. As shown in Fig. 2(b-c), illumination of the whole device area between the two electrodes significantly influenced the device response with increased device current. The data show that the device exhibits sensitivity across a wide wavelength range, from the visible to the near-infrared region. The photocurrent, defined as the difference between the laser on and off currents ($I_{Ph} = I_{laser\ on} - I_{laser\ off}$), was found to increase with the increase of bias voltage (Fig. 2(c)) within the visible-to-

NIR wavelengths, as expected. Consistent photocurrent and dark current levels across photo-switching cycles (i.e., laser on and off cycles) confirm a stable photo response to incident light.

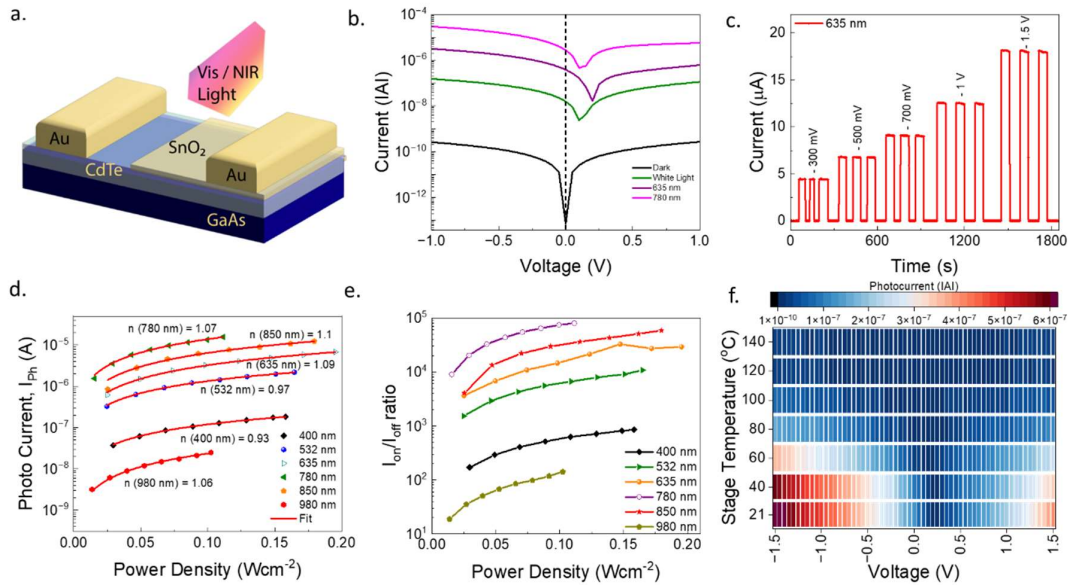

Fig. 2: (a) Schematic of the device structure, (b) quasi-static I-V characteristics upon illumination with different laser and white light sources, and (c) Light-induced increase of device current as a function of voltage bias with 60 s light ON/OFF time at an illumination with a 635 nm laser, (d-e) Photocurrent and $I_{on/off}$ ratio as a function of wavelength and power density, respectively, and (f) temperature-dependent photocurrents (measured with a laser of 850 nm wavelength and power density ~ 2.63 Wcm$^{-2}$) when the device was heated at elevated stage temperatures.

Figs. 2(d-e) show the light-intensity-dependent photo response of the device under varying laser wavelengths and power densities. The photocurrent increases with laser power (Fig. 2(d)), following a power-law relationship, $I_{Ph} \propto P^n$, where 'P' is the power density and 'n' is the power index. Across all tested wavelengths, 'n' deviates slightly from unity, ranging from 0.93 at 400 nm to 1.1 at 850 nm. This deviation is attributed to non-radiative carrier recombination mediated by defect states at shorter wavelengths (400 nm and 532 nm), as described in previous studies [42, 43]. At longer wavelengths (635, 780, 850, and 980 nm), the observed deviation is likely due to photothermal heating [44], which facilitates carrier generation through electron emission from trap states under increased photon flux density, a parameter directly proportional to the wavelength.

Importantly, the device demonstrated a high $I_{on}/I_{off}$ ratio, exhibiting a nearly 10$^5$-fold increase in current upon illumination with a 780 nm laser at a power density of 100 mWcm$^{-2}$ (Fig. 2(e)), clearly identifying its potential for high-sensitivity/low-noise device applications. The device reliability was further tested by varying the stage temperature and recording the photocurrent by illuminating the device with an 850 nm laser, as shown in Fig. 2(f). The figure demonstrates that the device maintained a distinctive photo response up to 80°C and continued to exhibit photo response at an elevated temperature of 140°C, albeit with a 30-fold reduction (measured at -1.5 V) compared to room temperature, indicating excellent performance in harsh environments. The bias polarity-dependent photocurrent is discussed in Section 2.3.

Responsivity and specific detectivity were further evaluated and benchmarked against a CdTe device without the SnO₂ layer. Measurements under six distinct laser illumination revealed that, for constant laser power density (100 mW cm⁻²), the device responsivity increased by nearly an order of magnitude, while the specific detectivity improved by two orders of magnitude (see Figs. 3(a-b)). The photocurrent data are given in the supporting information (Fig. S3).

Device response was also assessed under broadband illumination (420 nm - 1000 nm) using an 888-W tungsten-halogen lamp with measurements taken at each wavelength at 5 nm intervals to investigate the detailed spectral photo response. The photocurrent data, presented in Fig. 3(c), indicates enhanced sensitivity in the heterostructure, with a 5 nm resolution in wavelength. It is evident that while the CdTe device demonstrated optimal sensitivity near its bandgap (820 nm), the heterostructure displayed that within a broader wavelength range, such as between 500 and 940 nm.

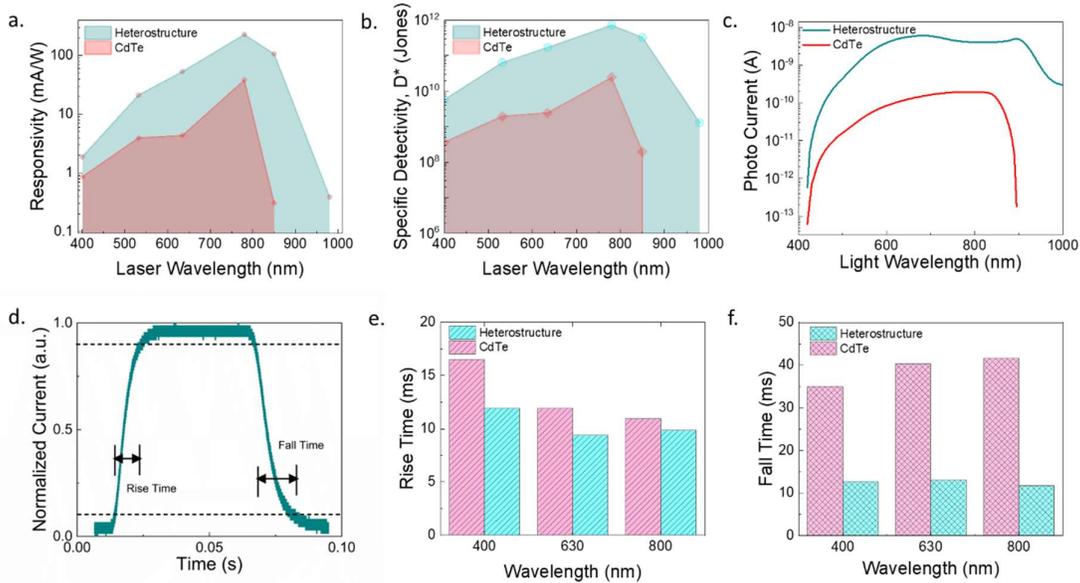

Fig. 3: Comparison of (a-b) responsivity and specific detectivity ($D^*$) for devices based on SnO₂/CdTe heterostructures and CdTe alone (The responsivity ($R$) of the photoconductor was defined as $R = I_{Ph}/P.S$, where P is the incident power intensity of the laser beam, and S is the active device area. $D^*$ calculated using the relation: $D^* = I_{Ph}.\sqrt{S}/(P.S.\sqrt{2q.I_{dark}})$ [45]. This measurement used lasers of 6 distinct wavelengths with 100 mW cm⁻² power density. (c) Photocurrent (measured at a bias voltage of -1 V) as a function of wavelength measured using a tungsten-halogen lamp by measuring photocurrent at each wavelength with a 5 nm interval (d) Dynamic photo response curve and (e-f) Extracted photo response times (rise and fall times, respectively) for SnO₂/CdTe heterostructures and CdTe-only devices when illuminated with different laser wavelengths.

To assess the response speed of the fabricated photodetectors under various illumination conditions, measurements of rise and fall times were performed, as summarized in Figs. 3(d-f). Fig. 3(d) illustrates a representative dynamic photo response curve, defining rise and fall times as the periods required for the current to increase from 10% to 90% of its peak value and decrease from 90% to 10%, respectively. These measurements offer critical insights into the dynamic performance of our device. $SnO_2$/CdTe heterostructures and CdTe-only devices exhibit characteristic response times within the millisecond range, with the heterostructure exhibiting a faster response irrespective of laser wavelengths, as summarised in Fig. 3(e-f). We also tested the photo response characteristics of devices with varying electrode gaps and employed CdTe substrates with different orientations. All the devices exhibited reliable photo responses, as shown in Fig. S4 (supporting information).

Table 1 summarizes the key performance metrics of our device, along with recently reported photodetectors based on 2D materials and CdTe. While direct comparisons are limited due to variations in device sizes, dimensions, choice of electrode metals, and operating conditions the fabricated photodetectors exhibit a notable enhancement in photocurrent and a broadband photo response, demonstrating its competitiveness.

Table 1: Comparison of Photodetection Performance

| Material | Wavelength range / Peak response | Bias Voltage | $I_{on}/I_{off}$ ratio | Laser Power (mW $cm^{-2}$) | Specific Detectivity (Jones) | Rise time/Fall time (ms) | Ref. |
|---|---|---|---|---|---|---|---|
| CdTe/CdS Core-shell nanostructure | 510 nm | -7.5 V | - | - | $4.5 \times 10^{11}$ | - | 46 |
| CdTe/$MoS_2$ | 405 nm | 0 V | - | 0.0739 | $5.84 \times 10^{11}$ | 0.044/0.13 | 47 |
| CdTe nanosheets | 473 nm | 1 V | 27 | 132.3 | $10^9$ | 18.4/14.7 | 48 |
| $Sb_2Se_3$ thin films | 940 nm | 5 V | 31 | - | $8.58 \times 10^{10}$ | 35/38 | 49 |
| $PdSe_2$ thin films | 730 nm | 0 V | - | - | $5.17 \times 10^{11}$ | 0.72/0.24 | 50 |
| GaSe/$VO_2$ | 405 nm | 3 V | 433 | 150.10 | $2.14 \times 10^{11}$ |  | 51 |
| $SnO_2$/CdTe | 400-980 nm | 0.5 V | ~$10^5$ @ 780 nm | 100 | $0.72 \times 10^{12}$ @ 780 nm | 9.8/11.7 @ 800 nm | This work* |

*The performance of the heterostructure and CdTe devices were compared using a 780 nm laser wavelength, as it was the closest available wavelength to the bandgap of CdTe due to the unavailability of 820 nm lasers in our lab.*

It is important to highlight that the reported figures of merit, including responsivity, specific detectivity, and the response time of our photodetector, are affected by the photoconductive uncoated CdTe and material inhomogeneity. These parameters could be further optimized by employing photolithographically defined selective area etching of $SnO_2$ to reduce the uncoated CdTe region and access the underlying CdTe layer for one of the electrode depositions. Additionally, scaling down the overall device area could enhance specific detectivity as it scales with area [52], and minimizing parasitic contributions from the measurement circuit may also improve device performance, such as response speed.

## 2.3 Mechanism of the enhanced photoresponse

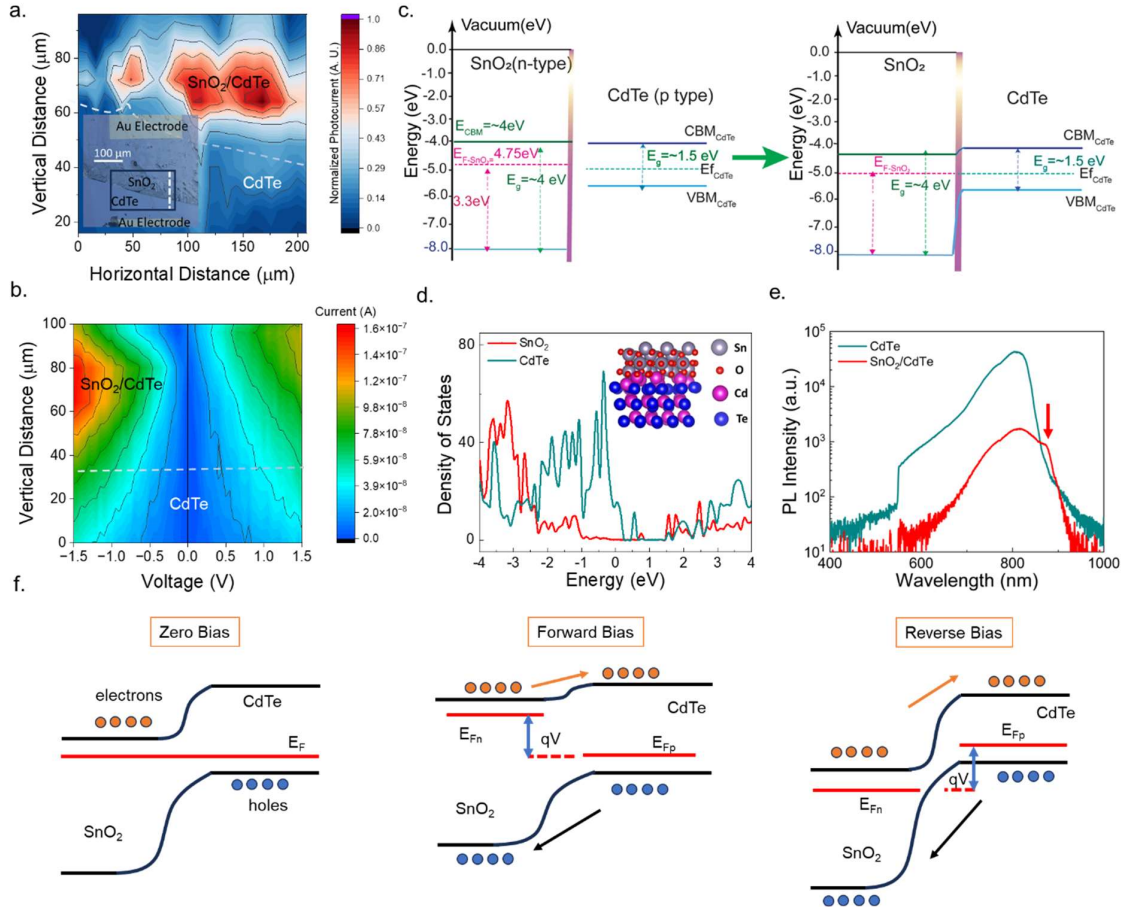

Fig. 4: (a) 2D photocurrent map of the selected device area marked by the rectangle in the inset showing an optical microscope image of the whole device area (Laser power was 0.6 μW at 700nm during photocurrent mapping); the white dashed line shows the boundary between device regions with $SnO_2$/CdTe and CdTe only and (b) photo-gated quasi-static I-V along the straight line shown inset of (a). (c) Schematic band diagrams of individual materials (experimental) and their heterostructure (expected), (d) Density of States as a function of energy calculated by DFT, (e) PL spectra of CdTe and $SnO_2$/CdTe heterostructure without applying any bias, and (f) schematics showing the electronic transport under different bias conditions (V is the magnitude of applied bias; the other notations have their usual meaning as described in a typical p-n junction).

To investigate the enhanced photo response of the heterostructure device, a photocurrent map was recorded by applying a 1V bias voltage and illuminating the device with a 700 nm laser. Fig. 4(a) shows the photocurrent map of the selected device area, identifying that both CdTe and the SnO$_2$/CdTe heterostructure contribute to the overall photocurrent response, with a dominant contribution from the heterostructure. A line scan was also conducted under illumination along the direction shown in Fig. 4(a) (indicated by the dashed straight line in the optical image), and quasi-static I-V curves were recorded along the line, which provides additional insight (Fig. 4(b)). In this configuration, the device exhibited forward-bias characteristics when a negative bias was applied to the electrode directly in contact with the SnO$_2$ layer and reverse-bias characteristics when a positive bias was applied to the same electrode. Under negative bias, photocurrent is generated across the entire scanned area, with the active region concentrating on the SnO$_2$/CdTe region. Importantly, the SnO$_2$/CdTe region exhibited more significant bias-polarity-dependent current asymmetry than the CdTe region, indicative of junction formation within this area (Fig. 4(b)). In this context, it is worth noting that the dark current characteristics also exhibit polarity-dependent current asymmetry (Fig. 2(b)), suggesting a complex current transport mechanism within the device structure. Despite this complexity, the photocurrent mapping distinctly reveals contributions from both the CdTe region and the SnO$_2$/CdTe heterostructure, with the latter demonstrating a dominant photocurrent response.

To elucidate the nature of the junction formed in the heterostructure, we further investigated the band structure, combining the results of different techniques, such as valence band XPS (VB-XPS), photoemission spectroscopy in the air (PESA), photoluminescence, and UV-visible spectroscopy. The experimental data from these measurements are provided in the supporting information (Figs. S5 and S6), and simplified band diagrams of the individual materials and that of the heterostructure are shown in Fig. 4(c). Our analysis identified that the liquid metal synthesis technique achieved n-type SnO$_2$ with an optical bandgap of ~ 4 eV and Fermi level at ~ -4.75 eV, consistent with the dominance of the SnO$_2$ phase in the nanosheet. In contrast, p-type CdTe with a bandgap of ~1.5 eV and Fermi level at ~ -5 eV was achieved by Molecular Beam Epitaxy, consistent with previous reports [53-55]. Consequently, a vertical p-n junction with a type-II band alignment [56] was formed when the two materials were hetero-integrated, notably without extrinsic doping. The p-type conductivity of undoped CdTe is related to intrinsic Cd vacancy [53-55], and the n-type nature of SnO$_2$ is due to embedded oxygen vacancy or defects introduced by the synthesis technique [57].

The band-alignment at the heterojunction strongly affects the charge transport across the interface and hence the overall optoelectronic response of the device and was further studied by density functional theory (DFT) calculations. Fig. 4(d) shows the DFT calculated electronic band structure, which is consistent with experimental findings, i.e., SnO$_2$ presents a large band gap of about 4 eV (VBM around -2.5 eV and CBM around 1.5 eV), and CdTe presents a smaller band gap of about 1.5 eV (VBM around the Fermi level and CBM around 1.5 eV). The valence band maximum of SnO$_2$ lies well below the Fermi level, while that of CdTe is positioned closer to it. In contrast, the conduction band minimum for CdTe and SnO$_2$ appear relatively similar above the Fermi level, consistent with the anticipated band alignment at the heterostructure interface (the partial density of states is given in the supporting information, Fig. S7).

Significantly, DFT analysis shows that several defect/hybridization states arise within the band gaps due to defect/hybridization states arising from the interface coupling, further indicated by the results of PL measurements, as shown in Fig. 4(e). Under excitation with a 532 nm laser, the $SnO_2$/CdTe structure exhibits a distinct emission peak at approximately 880 nm (as indicated by the arrow), alongside the dominant 820 nm peak associated with the direct bandgap energy of CdTe (1.5 eV). Electron-hole pairs generated at the conduction band-valence band edges could relax into these states at the interface, thus lowering their energy before recombination, which can enhance the broadband photocurrent.

The above analyses offer valuable insights into the enhanced photo response of the heterostructure device, which can be elucidated through the energy band diagrams presented in Fig. 4(f). Under zero-bias condition, most photogenerated electron-hole pairs are created within the CdTe layer, benefiting from its optimal bandgap. The built-in electric field at the CdTe/$SnO_2$ heterojunction effectively drives electrons toward the $SnO_2$ layer and holes toward the CdTe layer, facilitating carrier separation and reducing recombination losses. When a negative bias is applied to the *n*-type $SnO_2$, the barrier height for electrons in the $SnO_2$ is lowered, allowing electrons to traverse the junction more readily, as shown in Fig. 4(f), enhancing current generation. Conversely, applying a positive bias to *n*-type $SnO_2$ increases the potential barrier, suppressing current generation. Additionally, the high electron mobility of $SnO_2$ supports efficient photogenerated carrier collection, while its wide bandgap prevents holes from reaching the electrode, ensuring unidirectional current flow and further improving the photocurrent. The type-II band alignment in the heterostructure broadens its absorption spectrum due to band bending, enabling the capture of photons over a wider wavelength range. This generates more charge carriers and significantly improves the photo response as evident from Fig. 3(a-b). It is also noteworthy to highlight the contribution of photocurrent generated by photogenerated carriers in the CdTe layer (uncovered by $SnO_2$) that diffuse to the heterojunction and are collected by the built-in electric field. These carriers contribute to both the photocurrent and the dark current, potentially reducing the device speed. Overall, the enhanced photo response of the CdTe/$SnO_2$ heterostructure is achieved through efficient charge transport, reduced recombination, and broadened spectral sensitivity. The distinct electrical conductivities, favorable band alignment, and high electron transfer efficiency of the $SnO_2$ layer facilitate these advantages.

## 3. Conclusions:

In summary, we demonstrated the integration of liquid metal-printed ultrathin $SnO_2$ nanosheets onto CdTe thin films and unlocked a broadband-sensitive photodetector with a response from the visible to the near-infrared spectra. The device exhibited efficient generation of photocurrents, achieving a maximum current on/off ratio of around $10^5$ under 100 mW $cm^{-2}$ power density of a 780 nm laser, a responsivity of 233 $mAW^{-1}$, and a specific detectivity of approximately $10^{12}$ Jones, nearly two orders of magnitude higher than that of devices fabricated with pure CdTe thin films. Additionally, the device generated photocurrents at elevated temperatures up to 140°C (with efficient photo response up to 80°C), highlighting its robust thermal stability. Compared to a pure CdTe device, which is mostly responsive to wavelengths near its band gap (820 nm), the heterostructure displayed excellent optical sensitivity within a broader wavelength range, such as between 500 and 940 nm. Furthermore, the heterostructure demonstrated faster response times than the CdTe-only device, independent of the illumination wavelength. The enhanced performance of the heterostructure was attributed to the formation of a *p-n* junction, which occurs without external doping, as confirmed by band structure analysis, DFT calculations, and photocurrent mapping. This study presents a reproducible and scalable fabrication approach for developing next-generation nanoscale photodetectors and optoelectronic devices tailored for broadband applications.

## 4. Experimental Details

**Synthesis of 2D SnO$_2$:** 2D SnO$_2$ layers were exfoliated from the oxidized surface of liquid tin under an ambient environment by following the liquid metal synthesis technique. At first, elemental tin (purity 99.99%) purchased from Roto Metal was melted at a substrate temperature of 280°C. The melted metal was compressed between two glass slides to eliminate pre-existing oxide layers and potential airborne contaminants. Following this preparation, the newly formed oxide skin was gently detached by touching the surface with a pre-heated substrate (Si, SiO$_2$, and CdTe grown on GaAs) at 280°C (see Fig. 1(a)). The prepared nanosheets were then post-annealed at 450 °C for 15 minutes under ambient air conditions to achieve the dominant SnO$_2$ phase [27]. The CdTe thin films were grown using the molecular beam epitaxy (MBE) technique, as described in our earlier studies [58, 59], and the structural, optical, and compositional properties are given in the supporting information (Fig. S2).

**Material Characterization:** The structural properties of the samples were characterized using a Rigaku SmartLab X-ray diffractometer equipped with Cu K$\alpha$ radiation at a wavelength of 1.5406 Å. The instrument was operated at 40 kV and 40 mA, with an angular scanning range of $10° \leq 2\theta \leq 90°$ to capture fundamental diffraction peaks. Layer thicknesses and compositional analysis were analyzed using Bruker Dimension Icon AFM operating under ScanAsyst-air mode and a Thermo Scientific monochromatic Al K$\alpha$ XPS spectrometer (h$\nu$ ≈1486.6 eV) equipped with a concentric hemispherical analyzer, respectively. The XPS data was processed using CasaXPS software. TEM images were taken using a JEOL JEM-2100F TEM operating at an acceleration voltage of 200 kV equipped with a Gatan OneView camera. Gatan Microscopy suite software was utilized to analyze the data. TEM samples were prepared by directly printing the tin oxide sheets onto thermally and mechanically robust silicon nitride membrane window grids (Si$_3$N$_4$ grids, Ted Pella, 21587-10). Raman spectra of the synthesized and annealed 2D SnO$_x$ were carried out using a Raman spectrometer (Horiba Scientific LabRAM HR Evolution) equipped with a 532 nm laser source and 1800 mm$^{-1}$ grating. PESA Riken Keiki AC-2 was utilized to study the band structure of different samples at a power of 500 nW.

**Device Fabrication and Testing:** The ultrathin SnO$_2$-based photodetectors were fabricated using a direct writing photolithography process. First, AZ 5214E photoresist was spin-coated onto 2D tin oxide sheets deposited on MBE-grown CdTe thin films. The wafers were soft-baked on a hot plate at 95$^0$ C for 90 seconds. The electrodes were patterned using a Maskless

Aligner (Heidelberg MLA150). The samples were developed in AZ-726 MIF developer after laser writing. An e-beam evaporator (PVD75 - Kurt J. Lesker) was then employed to deposit Cr/Au (10/100 nm) electrodes, followed by a lift-off process defining a 250 μm × 250 μm active device area.

Electrical and optoelectronic measurements were performed using a Keysight B2902A Precision source unit, a Keithley 4200SCS semiconductor parameter analyzer, and an Agilent 2912A source meter. All the measurements were performed in the air, and biases were applied to the electrode deposited directly on top of 2D $SnO_2$ while the electrode deposited directly on top of CdTe was grounded. To assess the effects of light exposure, devices were subjected to laser illumination from sources with wavelengths of 400, 532, 630, 635, 700, 780, 850, and 980 nm, as well as to broadband light from a broadband 888-Watt Tungsten Halogen Lamp (Newport Model 66886). A SpectraPro 2300i monochromator was used to control the wavelength of incident light. The laser illumination covered the entire device area during the measurement, with the spot diameter around 250 μm. The photocurrent mapping was performed using a WhiteLase SC400 High-Power Supercontinuum laser with a circular spot diameter of 10 μm. A WiTec alpha300 S scanning near optical microscope (SNOM) was used to locate the laser spot on the device and adjust the three-axial stage moving step. Illumination power was calibrated using a Thorlabs power meter, and the photo response dynamics were recorded using a WaveAce 102, 60 MHz Oscilloscope.

**DFT Calculations:**
Density functional theory (DFT) calculations of the band structure were performed using the Projector Augmented Wave (PAW) method [60, 61] as implemented in the Vienna Ab initio Simulation Package (VASP) [62, 63]. The calculations were completed with a plane-wave cut-off energy of 500 eV and a single Gamma k-point mesh. The electronic self-consistent calculation was converged to $1×10^{-5}$ eV, and ionic relaxation steps were performed using the conjugate-gradient method (IBRION=2) and continued until the total force on each atom dropped below a tolerance of –0.05 eV/Å. The generalized gradient approximation (GGA) of the Perdew-Burke-Ernzerhof (PBE) approach [64] was used. Initially, the relaxation of CdTe (space group: $\bar{F}43m$ (Cubic); lattice parameter, a: 6.6237 Å) and $SnO_2$ (space group: $P4_2/mnm$ (Tetragonal); lattice parameters, a: 4.8242 Å and c: 3.2404 Å) unit cells were conducted, followed by the optimization of the slabs of the $CdTe/SnO_2$ interface. The joint interface slab model of CdTe and $SnO_2$ was developed based on experimental observation with the oxidized CdTe surface. The band alignment investigations were further calculated using the hybrid functionals approach of Heyd-Scuseria-Ernzerhof (HSE) using the relaxed structure obtained from the PBE approach [65].

**Acknowledgments**
S.K.N. acknowledges the salary and research support from the Forrest Research Foundation in the form of the Forrest Prospect Fellowship, ARC DP220101532, and ACNS grant proposal number 17399, respectively. N.S. acknowledges the support from the University of Melbourne in the form of a McKenzie postdoctoral fellowship. The authors acknowledge the support from TMOS CE200100010 grant and thank Dr. Renjie Gu, Songqing Zhang, Dr. Xiao Sun, and Dr.

Mark Lockrey for assisting with CdTe growth, characterization, and setting up an electrical testing unit.

**Author Contributions:**
**Shimul Kanti Nath** worked on conceptualization and project design, implementation, material characterization (XRD, SEM, PL), electrical measurements, photodetector measurements, photocurrent mapping, data analysis and interpretation, liaising with collaborators, and manuscript drafting.
**Nitu Syed** worked on project design, liquid metal printing, device fabrication, band structure investigation, Raman and AFM measurements, and partial manuscript drafting
**Wenwu Pan** worked on the growth and characterization of CdTe thin film.
**Yang Yu and Dawei Liu** performed light-controlled electrical measurements, including responsivity, photocurrent mapping, and photo response time measurements.
**Jodie Yuwono and Priyank Kumar** performed DFT calculations.
**Yan Zhu** performed PL measurements.
**David L. Cortie** performed XRD measurements of the heterostructure.
**Chung K. Nguyen** performed XPS measurements.
**Michael P. Nielsen, Lan Fu, Ann Roberts, Lorenzo Faraone, Nicholas J. Ekins-Daukes, and Wen Lei** supervised different portions of this study, provided research support when required, and aided in data interpretation.

**References**

1. Wang, J.; Han, J.; Chen, X.; Wang, X., Design strategies for two-dimensional material photodetectors to enhance device performance. *InfoMat* **2019,** *1* (1), 33-53.
2. Baugher, B. W.; Churchill, H. O.; Yang, Y.; Jarillo-Herrero, P., Optoelectronic devices based on electrically tunable *p–n* diodes in a monolayer dichalcogenide. *Nat. Nanotechnol.* **2014,** *9* (4), 262-267.
3. Das, S.; Sebastian, A.; Pop, E.; McClellan, C. J.; Franklin, A. D.; Grasser, T.; Knobloch, T.; Illarionov, Y.; Penumatcha, A. V.; Appenzeller, J., Transistors based on two-dimensional materials for future integrated circuits. *Nat. Electron.* **2021,** *4* (11), 786-799.
4. Koppens, F.; Mueller, T.; Avouris, P.; Ferrari, A.; Vitiello, M.; Polini, M., Photodetectors based on graphene, other two-dimensional materials and hybrid systems. *Nat. Nanotechnol.* **2014,** *9* (10), 780-793.
5. Ahmad, W.; Liu, J.; Jiang, J.; Hao, Q.; Wu, D.; Ke, Y.; Gan, H.; Laxmi, V.; Ouyang, Z.; Ouyang, F., Strong Interlayer transition in few-layer InSe/PdSe$_2$ van der Waals heterostructure for near-infrared photodetection. *Adv. Funct. Mater.* **2021,** *31* (43), 2104143.
6. Cho, S. W.; Jo, C.; Kim, Y.-H.; Park, S. K., Progress of materials and devices for neuromorphic vision sensors. *Nanomicro Lett.* **2022,** *14* (1), 203.
7. Ahmed, T.; Tahir, M.; Low, M. X.; Ren, Y.; Tawfik, S. A.; Mayes, E. L.; Kuriakose, S.; Nawaz, S.; Spencer, M. J.; Chen, H., Fully Light-controlled memory and neuromorphic computation in layered black phosphorus. *Adv. Mater.* **2021,** *33* (10), 2004207.
8. Li, X.; Zhu, J.; Wei, B., Hybrid nanostructures of metal/two-dimensional nanomaterials for plasmon-enhanced applications. *Chem. Soc. Rev.* **2016,** *45* (11), 3145-3187.

9. Lee, E.; Yoo, H., Self-powered sensors: new opportunities and challenges from two-dimensional nanomaterials. *Molecules* **2021,** *26* (16), 5056.
10. Yang, J.; Li, J.; Bahrami, A.; Nasiri, N.; Lehmann, S.; Cichocka, M. O.; Mukherjee, S.; Nielsch, K., Wafer-scale growth of $Sb_2Te_3$ films via low-temperature atomic layer deposition for self-powered photodetectors. *ACS Appl. Mater. Interfaces.* **2022,** *14* (48), 54034-54043.
11. Vashishtha, P.; Prajapat, P.; Sharma, A.; Singh, P.; Walia, S.; Gupta, G., Self-driven UVC–NIR broadband photodetector with high-temperature reliability based on a coco palm-like $MoS_2$/GaN heterostructure. *ACS Appl. Electron. Mater.* **2023,** *5* (3), 1891-1902.
12. Liu, J.; Hao, Q.; Gan, H.; Li, P.; Li, B.; Tu, Y.; Zhu, J.; Qi, D.; Chai, Y.; Zhang, W., Selectively modulated photoresponse in Type-I heterojunction for ultrasensitive self-powered photodetectors. *Laser Photonics Rev.* **2022,** *16* (11), 2200338.
13. Zeng, L. H.; Lin, S. H.; Li, Z. J.; Zhang, Z. X.; Zhang, T. F.; Xie, C.; Mak, C. H.; Chai, Y.; Lau, S. P.; Luo, L. B., Fast, self-driven, air-stable, and broadband photodetector based on vertically aligned $PtSe_2$/GaAs heterojunction. *Adv. Funct. Mater.* **2018,** *28* (16), 1705970.
14. Cui, T.; Lv, R.; Huang, Z.-H.; Chen, S.; Zhang, Z.; Gan, X.; Jia, Y.; Li, X.; Wang, K.; Wu, D., Enhanced efficiency of graphene/silicon heterojunction solar cells by molecular doping. *J. Mater. Chem. A* **2013,** *1* (18), 5736-5740.
15. Dhyani, V.; Das, S., High-speed scalable silicon-$MoS_2$ PN heterojunction photodetectors. *Sci. Rep.* **2017,** *7* (1), 1-9.
16. Shin, G. H.; Park, J.; Lee, K. J.; Lee, G.-B.; Jeon, H. B.; Choi, Y.-K.; Yu, K.; Choi, S.-Y.; interfaces, Si–$MoS_2$ vertical heterojunction for a photodetector with high responsivity and low noise equivalent power. *ACS Appl. Mater. Interfaces.* **2019,** *11* (7), 7626-7634.
17. Li, J.; Xi, X.; Lin, S.; Ma, Z.; Li, X.; Zhao, L., Ultrahigh sensitivity graphene/nanoporous GaN ultraviolet photodetectors. *ACS Appl. Mater. Interfaces.* **2020,** *12* (10), 11965-11971.
18. Zhao, S.; Zhang, J.; Fu, L., Liquid metals: a novel possibility of fabricating 2D metal oxides. *Adv. Mater.* **2021,** *33* (9), 2005544.
19. Daeneke, T.; Atkin, P.; Orrell-Trigg, R.; Zavabeti, A.; Ahmed, T.; Walia, S.; Liu, M.; Tachibana, Y.; Javaid, M.; Greentree, A. D.; Russo, S. P.; Kaner, R. B.; Kalantar-Zadeh, K., Wafer-scale synthesis of semiconducting SnO monolayers from interfacial oxide layers of metallic liquid tin. *ACS Nano* **2017,** *11* (11), 10974-10983.
20. Nguyen, C. K.; Low, M. X.; Zavabeti, A.; Murdoch, B. J.; Guo, X.; Aukarasereenont, P.; Mazumder, A.; Dubey, A.; Jannat, A.; Rahman, M. A.; Chiang, K.; Truong, V. K.; Bao, L.; McConville, C. F.; Walia, S.; Daeneke, T.; Syed, N., Atomically thin antimony-doped indium oxide nanosheets for optoelectronics. *Adv. Optical Mater.* **2022,** *20* (10), 2200925.
21. Aukarasereenont, P.; Goff, A.; Nguyen, C. K.; McConville, C. F.; Elbourne, A.; Zavabeti, A.; Daeneke, T., Liquid metals: an ideal platform for the synthesis of two-dimensional materials. *Chem. Soc. Rev.* **2022,** *51* (4), 1253-1276.
22. Zavabeti, A.; Ou, J. Z.; Carey, B. J.; Syed, N.; Orrell-Trigg, R.; Mayes, E. L.; Xu, C.; Kavehei, O.; O'Mullane, A. P.; Kaner, R. B., A liquid metal reaction environment for the room-temperature synthesis of atomically thin metal oxides. *Science* **2017,** *358* (6361), 332-335.
23. Du, B.; He, K.; Tian, G.; Che, X.; Song, L., Robust electron transport layer of $SnO_2$ for efficient perovskite solar cells: recent advances and perspectives. *J. Mater. Chem. C* **2023**.
24. Kumar, M.; Saravanan, A.; Joshi, S. A.; Chen, S.-C.; Huang, B.-R.; Sun, H., High-performance self-powered UV photodetectors using $SnO_2$ thin film by reactive magnetron sputtering. *Sens. Actuators A: Phys.* **2024,** *373*, 115441.

25. Zhang, Y.; Xie, Q.; Lyu, L.; Zou, J.; Lu, Z., Nanostructured $SnO_2$ thin films based on a convenient chemical deposition for sensitive detection of ethanol. *ACS Omega* **2024,** *9* (14), 16055-16062.
26. Kanth, S.; Chikara, A. K.; Choudhury, S.; Betty, C. In *Portable $H_2S$ gas detection and alert system prototype using highly sensitive nanocrystalline $SnO_2$ thin films: Sensor for environmental monitoring*, 2023 IEEE Applied Sensing Conference (APSCON), IEEE: 2023; pp 1-3.
27. Nguyen, C. K.; Taylor, P. D.; Zavabeti, A.; Alluhaybi, H.; Almalki, S.; Guo, X.; Irfan, M.; Kobaisi, M. A.; Ippolito, S. J.; Spencer, M. J., Instant-in-air liquid metal printed ultrathin tin oxide for high-performance ammonia sensors. *Adv. Funct. Mater.* **2024,** *34* (31), 2309342.
28. Batzill, M.; Diebold, U., The surface and materials science of tin oxide. *Prog. Surf. Sci.* **2005,** *79* (2-4), 47-154.
29. Karmaoui, M.; Jorge, A. B.; McMillan, P. F.; Aliev, A. E.; Pullar, R. C.; Labrincha, J. o. A. n.; Tobaldi, D. M., One-step synthesis, structure, and band gap properties of $SnO_2$ nanoparticles made by a low temperature nonaqueous sol–gel technique. *ACS Omega* **2018,** *3* (10), 13227-13238.
30. Godinho, K. G.; Walsh, A.; Watson, G. W., Energetic and electronic structure analysis of intrinsic defects in $SnO_2$. *J. Phys. Chem. C* **2009,** *113* (1), 439-448.
31. Hu, Y.; Hwang, J.; Lee, Y.; Conlin, P.; Schlom, D. G.; Datta, S.; Cho, K., First principles calculations of intrinsic mobilities in tin-based oxide semiconductors SnO, $SnO_2$, and $Ta_2SnO_6$. *J. Appl. Phys.* **2019,** *126* (18).
32. Birkmire, R. W.; McCandless, B. E., CdTe thin film technology: Leading thin film PV into the future. *Curr. Opin. Solid State Mater. Sci.* **2010,** *14* (6), 139-142.
33. Nagaura, T.; Phan, H. P.; Malgras, V.; Pham, T. A.; Lim, H.; Ashok, A.; Kim, J.; You, J.; Nguyen, N. T.; Na, J., Universal electrochemical synthesis of mesoporous chalcogenide semiconductors: Mesoporous CdSe and CdTe thin films for optoelectronic applications. *Angew. Chem.* **2021,** *133* (17), 9746-9751.
34. Lian, Q.; Zhu, X.; Wang, X.; Bai, W.; Yang, J.; Zhang, Y.; Qi, R.; Huang, R.; Hu, W.; Tang, X., Ultrahigh-Detectivity Photodetectors with Van der Waals Epitaxial CdTe Single-Crystalline Films. *Small* **2019,** *15* (17), 1900236.
35. Xie, X.; Kwok, S.-Y.; Lu, Z.; Liu, Y.; Cao, Y.; Luo, L.; Zapien, J. A.; Bello, I.; Lee, C.-S.; Lee, S.-T., Visible–NIR photodetectors based on CdTe nanoribbons. *Nanoscale* **2012,** *4* (9), 2914-2919.
36. Sharan, A.; Nardone, M.; Krasikov, D.; Singh, N.; Lany, S., Atomically thin interlayer phase from first principles enables defect-free incommensurate $SnO_2$/CdTe interface. *Appl. Phys. Rev.* **2022,** *9* (4).
37. McGott, D. L.; Perkins, C. L.; Reese, M. O., Thermomechanical cleave of polycrystalline CdTe solar cells and its applications: a review. *Sol. RRL* **2023,** *7* (15), 2300074.
38. Syed, N.; Nguyen, C. K.; Zavabeti, A.; Low, M. X.; Wei, X.; Krishnamurthi, V.; Irfan, M.; Lee, W. S.; Duong, N. M.; Nguyen, A. T., Vacuum-free liquid-metal-printed 2D semiconducting tin dioxide: the effect of annealing. *ACS Appl. Electron. Mater.* **1842**.
39. Nose, K.; Suzuki, A.; Oda, N.; Kamiko, M.; Mitsuda, Y., Oxidation of SnO to $SnO_2$ thin films in boiling water at atmospheric pressure. *Appl. Phys. Lett.* **2014,** *104* (9).
40. Zarrinkhameh, M.; Zendehnam, A.; Hosseini, S.; Robatmili, N.; Arabzadegan, M., Effect of oxidation and annealing temperature on optical and structural properties of $SnO_2$. *Bull. Mater. Sci.* **2014,** *37*, 533-539.
41. Liu, A.; Kim, Y.-S.; Kim, M. G.; Reo, Y.; Zou, T.; Choi, T.; Bai, S.; Zhu, H.; Noh, Y.-Y., Selenium alloyed tellurium oxide for amorphous p-channel transistors. *Nature* **2024**, 1-3.

42. Luo, H.; Wang, H.; Zhang, S.; Liu, J.; Ren, Y.; Yuan, C.; Lei, W., Controlled growth of high-quality β-Ag$_2$Se nanowires and their applications in near-infrared photodetection. *J. Alloys Compd.* **2024,** *1008*, 176819.
43. Zhao, Q.; Wang, W.; Carrascoso-Plana, F.; Jie, W.; Wang, T.; Castellanos-Gomez, A.; Frisenda, R., The role of traps in the photocurrent generation mechanism in thin InSe photodetectors. *Mater. Horiz.* **2020,** *7* (1), 252-262.
44. Tian, J.; Chen, L.; Qiao, R.; Xiong, K.; Zhang, W.; Mao, Y.; Li, H.; Li, J., Photothermal-assist enhanced high-performance self-powered photodetector with bioinspired temperature-autoregulation by passive radiative balance. *Nano Energy* **2021,** *79*, 105435.
45. Wang, H.; Zhang, Z.; Luo, H.; Zhang, S.; Pan, W.; Liu, J.; Ren, Y.; Lei, W., A case study of 2D Bi$_2$O$_2$Se nanoplate near-infrared photodetectors from the perspective of practical applications. *Adv. Opt. Mater.* **2024,** *30* (12), 2401404.
46. Abdalaah, Y. K.; Mahmood, O. A.; Shaker, S. S.; Ismail, R. A., Preparation of a nanostructured CdTe@ CdS core–shell/Si photodetector by two-step laser ablation in liquid. *J. Opt.* **2024**, 1-12.
47. Yang, S.; Liu, Y.; Wu, Y.; Guo, F.; Zhang, M.; Zhu, X.; Xu, R.; Hao, L., High-performance flexible photodetectors based on CdTe/MoS$_2$ heterojunction. *Nanoscale* **2024,** *16* (29), 13932-13937.
48. Cheng, R.; Wen, Y.; Yin, L.; Wang, F.; Wang, F.; Liu, K.; Shifa, T. A.; Li, J.; Jiang, C.; Wang, Z., Ultrathin single-crystalline CdTe nanosheets realized via van der Waals epitaxy. *Adv. Mater.* **2017,** *29* (35), 1703122.
49. Wen, X.; Lu, Z.; Valdman, L.; Wang, G.-C.; Washington, M.; Lu, T.-M., High-crystallinity epitaxial Sb$_2$Se$_3$ thin films on mica for flexible near-infrared photodetectors. *ACS Appl. Mater. Interfaces.* **2020,** *12* (31), 35222-35231.
50. Zhang, X.; Dai, M.; Deng, W.; Zhang, Y.; Wang, Q. J., A broadband, self-powered, and polarization-sensitive PdSe$_2$ photodetector based on asymmetric van der Waals contacts. *Nanophotonics* **2023,** *12* (3), 607-618.
51. Zhou, J.; Xie, M.; Ji, H.; Cui, A.; Ye, Y.; Jiang, K.; Shang, L.; Zhang, J.; Hu, Z.; Chu, J., Mixed-dimensional van der Waals heterostructure photodetector. *ACS Appl. Mater. Interfaces.* **2020,** *12* (16), 18674-18682.
52. Pan, W.; Liu, J.; Zhang, Z.; Gu, R.; Suvorova, A.; Gain, S.; Wang, H.; Li, Z.; Fu, L.; Faraone, L., Large area van der Waals epitaxy of II–VI CdSe thin films for flexible optoelectronics and full-color imaging. *Nano Res.* **2022,** *15* (1), 368-376.
53. Emanuelsson, P.; Omling, P.; Meyer, B.; Wienecke, M.; Schenk, M., Identification of the cadmium vacancy in CdTe by electron paramagnetic resonance. *Phys. Rev. B* **1993,** *47* (23), 15578.
54. Kavanagh, S. R.; Walsh, A.; Scanlon, D. O., Rapid recombination by cadmium vacancies in CdTe. *ACS Energy Lett.* **2021,** *6* (4), 1392-1398.
55. Lindström, A.; Mirbt, S.; Sanyal, B.; Klintenberg, M., High resistivity in undoped CdTe: carrier compensation of Te antisites and Cd vacancies. *J. Phys. D: Appl. Phys.* **2015,** *49* (3), 035101.
56. Yang, J.; Ye, Y.; Yu, R.; Yang, H.; Qiao, H.; Huang, Z.; Qi, X., Enhanced UV photoresponse performances of TiO$_2$/Bi$_2$Se$_3$ heterostructure-based photoelectrochemical photodetector. *Phys. Status Solidi A*, 2400522.
57. Dangi, R.; Basnet, B.; Pandey, M.; Bhusal, S.; Budhathoki, B.; Parajuli, K.; Tiwari, S. K.; Kafle, B. P., Effect of oxygen vacancy on the crystallinity and optical band gap in tin oxide thin film. *Energies* **2023,** *16* (6), 2653.
58. Pan, W.; Ma, S.; Sun, X.; Nath, S. K.; Zhang, S.; Gu, R.; Zhang, Z.; Faraone, L.; Lei, W., Structural properties of MBE-grown CdTe (133) B buffer layers on GaAs (211) B substrates with CdZnTe/CdTe superlattice-based dislocation filtering layers. *J. Appl. Phys.* **2023,** *133* (18).
59.     Pan, W.; Nath, S. K.; Ma, S.; Gu, R.; Zhang, Z.; Fu, L.; Faraone, L.; Lei, W., Non-invasive and non-destructive characterization of MBE-grown CdZnTe/CdTe superlattice-based dislocation filtering layers. *J. Appl. Phys.* **2022,** *131* (20).
60.     Blöchl, P. E., Projector augmented-wave method. *Phys. Rev. B* **1994,** *50* (24), 17953.
61.     Kresse, G.; Joubert, D., From ultrasoft pseudopotentials to the projector augmented-wave method. *Phys. Rev. B* **1999,** *59* (3), 1758.
62.     Kresse, G.; Furthmüller, J., Efficiency of ab-initio total energy calculations for metals and semiconductors using a plane-wave basis set. *Comput. Mater. Sci.* **1996,** *6* (1), 15-50.
63.     Kresse, G.; Furthmüller, J., Efficient iterative schemes for ab initio total-energy calculations using a plane-wave basis set. *Phys. Rev. B* **1996,** *54* (16), 11169.
64.     Perdew, J. P.; Burke, K.; Ernzerhof, M., Generalized gradient approximation made simple. *Phys. Rev. Lett.* **1996,** *77* (18), 3865.
65.     Heyd, J.; Scuseria, G. E.; Ernzerhof, M., Hybrid functionals based on a screened Coulomb potential. *J. Chem. Phys.* **2003,** *118* (18), 8207-8215.

# Supporting Information

**Liquid Metal-Exfoliated SnO$_2$-Based Mixed-dimensional Heterostructures for Visible-to-Near-Infrared Photodetection**

Shimul Kanti Nath[1,2]*, Nitu Syed[3,4]*, Wenwu Pan[2], Yang Yu[5], Dawei Liu[5,6], Michael P. Nielsen[1], Jodie Yuwono[7], Priyank Kumar[8], Yan Zhu[1], David L. Cortie[9], Chung K. Nguyen[4], Lan Fu[5], Ann Roberts[3], Lorenzo Faraone[2], Nicholas J. Ekins-Daukes[1], and Wen Lei[2]

[1]School of Photovoltaic and Renewable Energy Engineering, University of New South Wales (UNSW Sydney), Kensington NSW 2052, Australia
[2]Australian Research Council Centre of Excellence for Transformative Meta-Optical Systems, Department of Electrical, Electronic and Computer Engineering, The University of Western Australia, 35 Stirling Highway, Perth, WA 6009, Australia
[3]Australian Research Council Centre of Excellence for Transformative Meta-Optical Systems, School of Physics, University of Melbourne, Victoria 3010, Australia
[4]School of Engineering, RMIT University, Melbourne, VIC, 3001 Australia
[5]Australian Research Council Centre of Excellence for Transformative Meta-Optical Systems, Department of Electronic Materials Engineering, Research School of Physics, The Australian National University, Canberra ACT 2601, Australia
[6]Institute of Applied Physics, Abbe Center of Photonics, Friedrich Schiller University Jena, Albert-Einstein-Straße 15, Jena, 07745, Germany
[7]School of Chemical Engineering, The University of Adelaide, Adelaide, SA 5005, Australia
[8]School of Chemical Engineering, University of New South Wales (UNSW Sydney), Kensington NSW 2052, Australia
[9]Australian Centre for Neutron Scattering, Australian Nuclear Science and Technology Organisation, Lucas Heights, NSW 2234, Australia
*Corresponding authors
  *Shimul_kanti.nath@unsw.edu.au
  *Nitu.syed@unimelb.edu.au

---

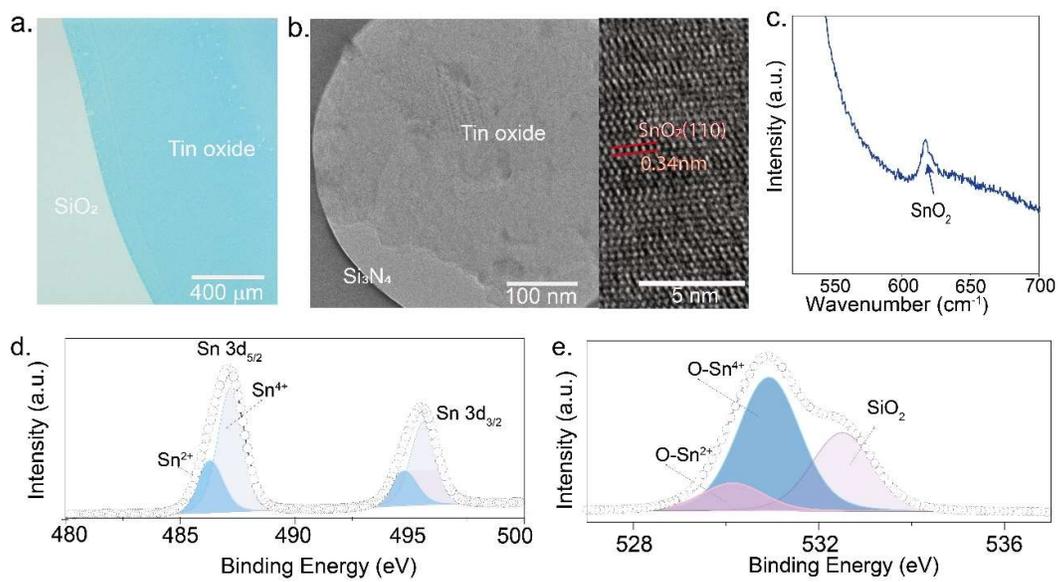

Fig. S1: Characterizations of liquid metal-derived tin oxide. (a) Optical image of a representative $SnO_x$ nanosheet. (b) TEM image of tin oxide nanosheets on holey silicon nitride support film. (c) Raman spectra of the nanosheets showing $SnO_2$ as the dominant phase. (d-e) XPS spectra of the Sn 3d and O 1s regions of tin oxide nanosheets prepared on 300 nm $SiO_2$/Si substrate.

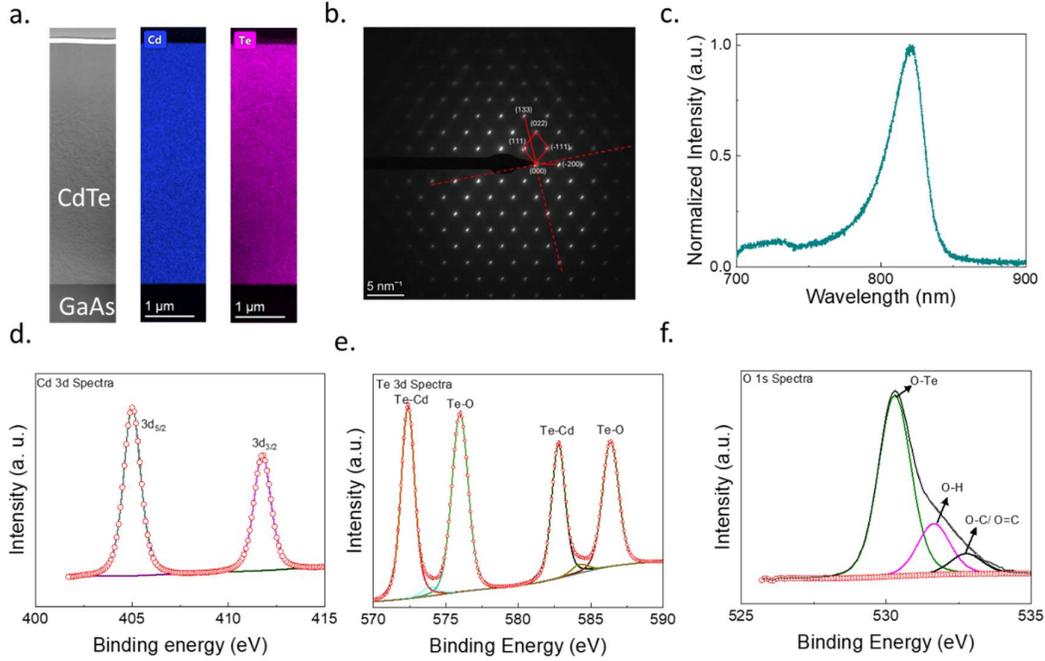

Fig. S2: Properties of MBE-grown CdTe substrate (a) Cross-sectional SEM and EDX maps showing homogeneous elemental distribution in the 133(B) CdTe. (b) Electron diffraction pattern (SAED) showing crystal orientation, (c) Photoluminescence spectra of as-grown CdTe indicating its optical bandgap at around 1.5 eV corresponding to the wavelength 820 nm, and (d) XPS spectra of CdTe showing the oxidation of Te at the surface.

**Thermodynamic calculations showing the prevalence of $SnO_2$:**

Using thermochemical data from Ref. [1], the change in Gibbs potential $\Delta G$ was determined. By considering the $\Delta G$ values of SnO, $SnO_2$, and $H_2O$ (-256.765 kJ/mol, -520 kJ/mol, and -237 kJ/mol, respectively), the following spontaneous reactions can be predicted:

$$Sn + O_2 = SnO_2 \ ; \Delta G = -520 \frac{kJ}{mol}$$

$$Sn + 2H_2O = SnO_2 + 2H_2 \ ; \Delta G = -45.72 \frac{kJ}{mol}$$

$$2SnO + O_2 = 2SnO_2 \ ; \Delta G = -263.235 \frac{kJ}{mol}$$

$$SnO + H_2O = SnO_2 + H_2 \ ; \Delta G = -26.094 \frac{kJ}{mol}$$

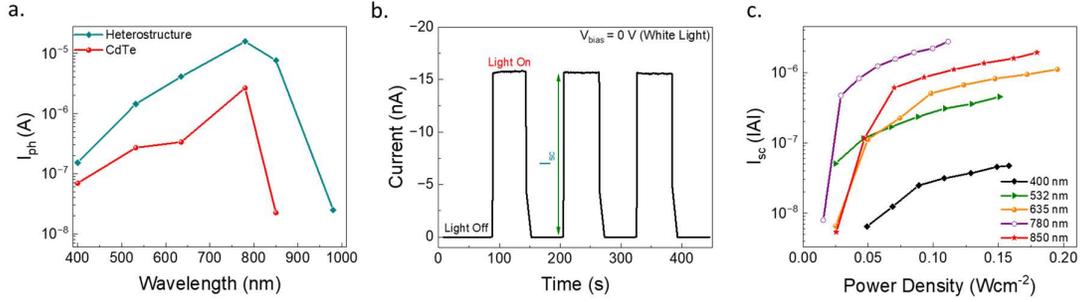

Fig. S3: (a) Photocurrent (Iph) measured at 0.5 V and with laser intensity 100 mW/cm$^2$ as a function of light wavelength. (b) Dynamic photo response measured at 0 V when the device is illuminated with a white light of arbitrary intensity, and (c) short circuit current ($I_{sc}$: photocurrent at zero bias) as a function of laser wavelength and power density.

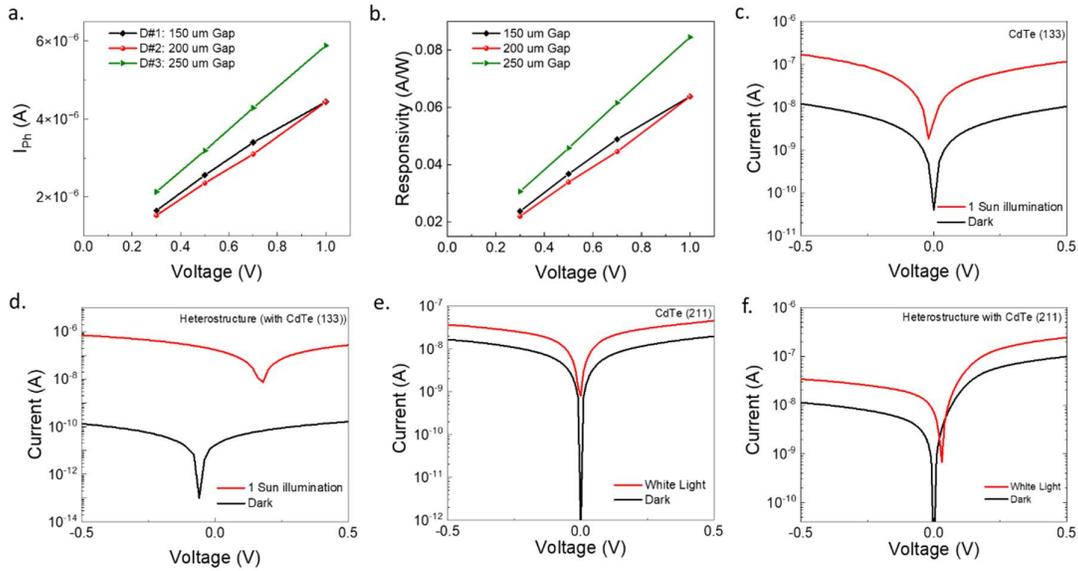

Fig. S4: (a) Photocurrent (Iph) and (b) Responsivity of 3 devices with different electrode gap distances when illuminated with a 780 nm laser (bias voltage = 0.5 V, power density 100 mW/cm$^2$). Note that the devices had variable $SnO_2$ area partially covering the gap schematically shown in Fig. 2(a). (c-d) Dark and light I-V of CdTe and $SnO_2$/CdTe under 1-Sun illumination with CdTe having 133 growth direction. (e-f) Dark and light I-V of CdTe and $SnO_2$/CdTe under white light of arbitrary intensity with CdTe having 211 growth direction.

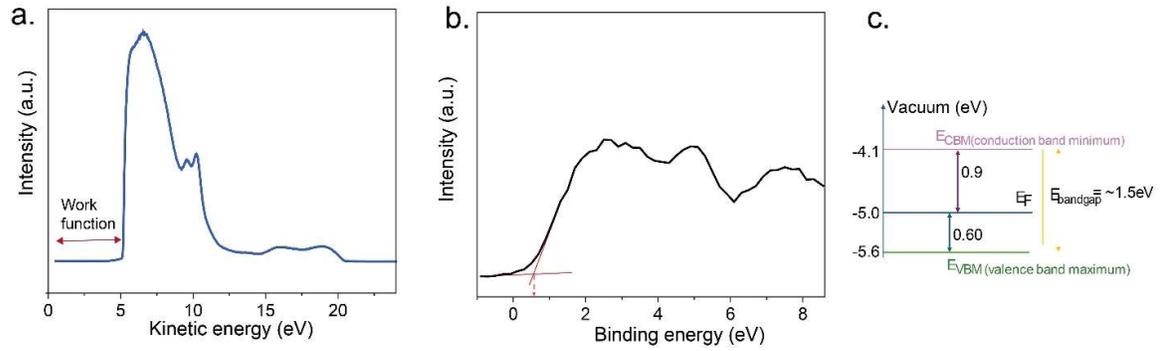

Fig S5: Band diagram characterizations of CdTe. (a) UPS analysis showing the work function of CdTe located at ~5 eV. (b) VBXPS analysis shows an energy difference of ~0.6 eV between the VBM and Fermi level of the CdTe. (c) proposed simplified band diagram of CdTe.

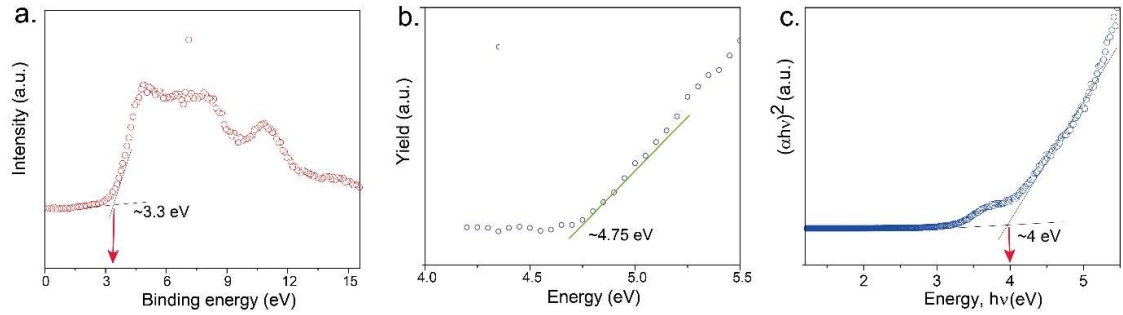

Fig. S6: Band diagram characterizations for the synthesized tin oxide. (a) VB XPS analysis of tin oxide shows an energy difference of 3.3 eV between the VBM and Fermi levels. (b) PESA spectra showing the fermi level is situated at 4.75 eV. (c) Tauc-plot obtained from the UV-vis data of the nanosheets where plot $\alpha$ and $h\nu$ represent the optical absorption coefficient and photon energy, respectively. Tauc plot indicates an optical bandgap of ~4 eV.

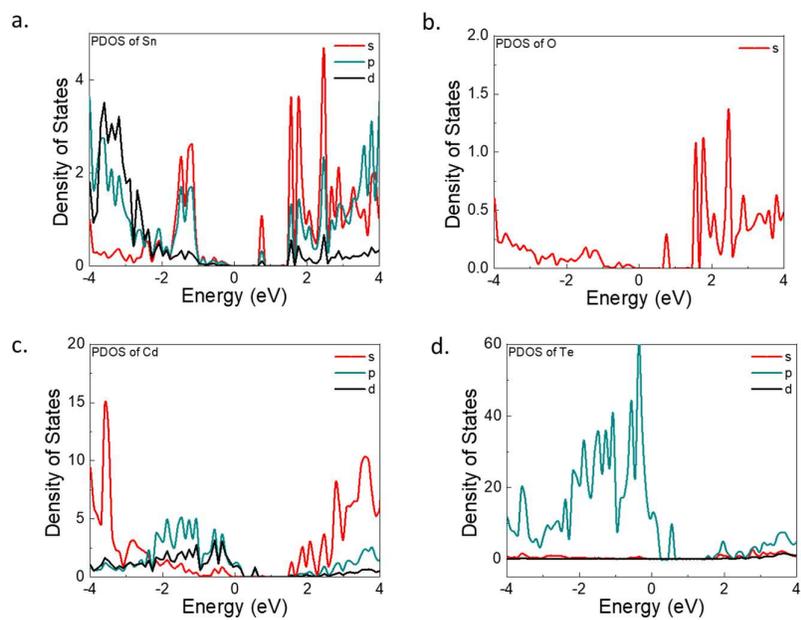

Fig. S7: Partial Density of States (PDOS) of various elements present in the heterostructure from DFT calculations

**References**

1. Ihsan, B., *Thermochemical data of pure substances.* and, 1995. **934**: p. 587.